# On Dimensional Analysis, Redundancy in set of fundamental quantities and Proposal of a New Set

**Mohd Abubakr,**
Microsoft R & D, Hyderabad,
Email: mohdabubakr@gmail.com

**ABSTRACT**

Inclusion of redundant fundamental quantities in SI system has resulted in lot of ambiguities and confusion in modern theories. The incompatibilities between the existing theories can possibly be due to incorrect assumption of fundamental quantities and existence of newer fundamental quantities. This paper is an attempt to eliminate the redundancy in the SI system using random mathematical mappings done through a computer program on SI system to New System. Each mathematical mapping is studied manually. Out of 1000 random mathematical mappings generated using a Computer Program, a mapping with "three fundamental quantities" is found to describe a non-redundant and intuitive set of Fundamental quantities. This paper also proposes the possible existence of a new fundamental quantity. A significant attempt is made to understand the advantages of new system over SI system. The relations between the set {Mass, length, time, current, and temperature} and new set of three fundamental quantities are calculated. The paper also describes the intuitive reasons favoring the new fundamental set of quantities.

1. INTRODUCTION

The idea of fundamental quantities of physics is nothing new to the scientific literature. Time and again there have been modifications in the notion of the primary constituents of the matter and Universe. As per the ancient Greek Philosophers, matter constituted of one single element. Later, it was modified to four fundamental elements i.e. fire, earth, water and air. Ancient Indians had theories with five fundamental elements i.e. Agni (fire), Pritvi (earth), jala (water), vayu (air) and akasha (Ether). Though scientifically this idea has been abandoned since a long time, nevertheless it was the stepping-stone about understanding the fundamental constituents of the Universe. Analyzing the reasons of attributing a certain physical quantity as fundamental reveals that such attribution is primarily dependent on the science of the existing era. If the researchers were unable to probe further into a quantity or were unable to split it into more fundamental gradients, they attributed such quantities as 'fundamental'.

The fundamental quantities that are currently being accepted by the scientific community are mass, time, length, current, temperature, luminous intensity and amount of substance also referred as SI units [9,10]. Amount of substance and luminous intensity are mere numbers and are continued as fundamental quantities only due to historic reasons. Ignoring luminous intensity and amount of substance from the further discussion, the present fundamental set consists of five constituents, namely mass, time, length, current and temperature. In order to reduce the numeric complexity of fundamental constants, most of the scientific literature is expressed in Natural Units such as Planck's units [11]. However, even the so-called Natural Units can dimensionally expressed in SI Units.

As our modern theories are built on the assumptions of fundamental quantities, any conflict in the choosing the fundamental quantities could result into chaos. If the fundamental quantities turn out to be redundant or incorrect, then the very foundation of any modern theory will fall trembling. With the multiple theories existing in modern day physics with conflicting results, it is noteworthy to doubt the correctness of existing set of fundamental quantities. In case, the existing set of fundamental quantities is wrong, then there are different possibilities such as
1. Redundancy in Fundamental quantities
2. Existence of new fundamental quantities
3. Redundancy in fundamental quantities and Existence of new fundamental quantities.

This paper proceeds with an assumption that "there exists an inherent redundancy in the existing fundamental quantities and existence of new fundamental quantities". If the assumption is wrong, then there should not exist a new set that can satisfy the existing scientific literature. The simplest ways to prove the correctness of the assumption is discover a new fundamental quantity and experimentally locate a redundancy in fundamental quantities. However, the method adopted to prove the assumption in this paper is rather different.

## 2. HUNT FOR FUNDAMENTAL SET

The assumption of existence of inherent redundancy and existence of new fundamental quantities can be interpreted as following, "there exists a new fundamental set with less than five fundamental quantities and Mass, Time, Length, current and temperature can be dimensionally expressed in terms of new fundamental quantities". In order to achieve this task a computer program is written that maps the existing five fundamental quantities into a new smaller set of fundamental quantities. The new fundamental set is purely mathematical. The algorithm used in the computer program has no intelligence and it merely generates different mappings. The only advantage of this computer program is it generates hundreds of mappings within a fraction of a second. The algorithm was run to produce more than 1000 different mappings.

Once the mappings are generated, each mapping was manually analyzed. While analyzing the mapping of fundamental quantities into new fundamental quantities, it is sensible to take the existing knowledge base (existing literature) into account. The new prediction can only be justifiable if it satisfies the all constraints of the existing knowledge base and also opens horizons for future research. If the assumption about the 'existing redundancy in the fundamental quantities and existence of new fundamental quantities' is incorrect, then there exists no new set that satisfies the existing knowledge base.

3. ALGORITHM

A program was written to map the five base quantities {M, L, T, K and A} into new three base quantities {X, Y and Z}. A simple three-step algorithm followed by the program is given below. Let B and N represent the set of base quantities {M, L, T, K and A} and {X, Y and Z} respectively. The algorithm is as follows

Step 1:
Assign the dimensional formula $X^aY^bZ^c$ to each base quantity of set B, where a, b and c are three random variables chosen differently for each base quantity.
Step 2:
Substitute the new dimensional formula of set B in the dimensional formula of the derived quantities such as force, energy etc.
Step 3:
Save all the dimensional formulae and different values of {a, b, c}.

The saved results in the Step 3 in the algorithm are analyzed manually. Note that the algorithm purely relies on the probability of the random combination of the N assigned to B. The above algorithm was implemented for 1000 random combinations of N for B. The next section of the paper puts light on the process that was carried out to implement the above algorithm.

4. PROGRAM OVERVIEW

A computer program was written to generate 3 random values {a, b and c} from the set R {-6.00, -5.50, -5.00, -4.50, -4.00, -3.50, -3.00, -2.50, -2.00, -1.50, -1.00, -0.50,

0.00, 0.50, 1.00, 1.50, 2.00, 2.50, 3.00, 3.50, 4.00, 4.50, 5.00, 5.50, 6.00}. Set R consists of total 25 values. Once the 3 random values are obtained, the new dimensional formula $X^a Y^b Z^c$ for a particular quantity from the Set B is created. Once all the quantities in the Set B are assigned a new dimensional formula, the corresponding values of {M, L, T, K and A} are substituted in the dimensional formulae of derived quantities such as velocity, force, energy, charge, permittivity, permeability, specific heat etc. In all, the program contains a database of 40 derived quantities. The resultant set of dimensional formulae of the derived quantities is taken out in a printed sheet and analysis is done manually. Presence of ambiguities in the dimensional formulae reflects that random combination of Set N assigned to Set B is redundant and wrong. Below is the example of random combination that resulted in the ambiguities.

5. **EXAMPLE FOR A RANDOM COMBINATION**

One of the random combinations generated by the program is given as below

$M = X^1 Y^0 Z^0$    for {a = 1, b = 0 and c = 0)
$L = X^0 Y^1 Z^0$    for {a = 0, b = 1 and c = 0)
$T = X^0 Y^0 Z^1$    for {a = 0, b = 0 and c = 1)
$K = X^1 Y^1 Z^0$    for {a = 1, b = 1 and c = 0)
$A = X^0 Y^1 Z^1$    for {a = 0, b = 1 and c = 1)

The new dimensional formulae obtained for {M, L, T, K and A} are substituted in the dimensional formulae of derived quantities. Consider the dimensional formulae of derived quantities obtained for the above substitutions,

Velocity = $[L^1 T^{-1}]$ = $[X^0 Y^1 Z^{-1}]$
Acceleration = $[L^1 T^{-2}]$ = $[X^0 Y^1 Z^{-2}]$
Force = $[M^1 L^1 T^{-2}]$ = $[X^1 Y^1 Z^{-2}]$
Energy = $[M^1 L^2 T^{-2}]$ = $[X^1 Y^2 Z^{-2}]$
Universal Gravitational Constant = $[M^{-1} L^3 T^{-2}]$ = $[X^{-1} Y^3 Z^{-2}]$
Charge = $[AT]$ = $[X^0 Y^1 Z^2]$
Voltage = $[M^1 L^2 T^{-3} A^{-1}]$ = $[X^1 Y^1 Z^{-4}]$
Magnetic Pole Strength = $[A^1 L^1]$ = $[X^0 Y^2 Z^1]$
Magnetic Moment = $[A^1 L^2]$ = $[X^0 Y^3 Z^1]$
Intensity of magnetization = $[A^1 L^{-1}]$ = $[X^0 Y^0 Z^1]$
Specific Heat = $[L^2 T^{-2} K^{-1}]$ = $[X^{-1} Y^1 Z^{-2}]$
Thermal Capacity = $[M^1 L^2 T^{-1} K^{-1}]$ = $[X^0 Y^1 Z^{-1}]$
Permeability = $[M^1 L^1 T^{-2} A^{-2}]$ = $[X^1 Y^{-1} Z^{-4}]$
Permittivity = $[M^{-1} L^{-3} T^4 A^2]$ = $[X^{-1} Y^{-1} Z^6]$
Impedance = $[M^1 L^2 T^{-3} I^{-2}]$ = $[X^1 Y^0 Z^{-5}]$

It can be seen that derived quantities have retained their uniqueness even after the 5 base quantities were transformed into 3 base quantities. Had their been no redundancy in the Set B {M, L, T, K and A}, the resultant set of formulae of derived quantities in the above example should have conflicted with each other. One of the challenging tasks here is to make the dimensional formulae for derived quantities more perceptive to practical observations. Hence the more number of random combinations were further probed to find the combinations that will have additional advantages over others. The next section of the paper gives the detailed analysis of a new dimensional system that is found to be most intuitively represents the practical observations out of 1000 mappings that were manually analyzed.

## 6. NEW DIMENSIONAL SET

Out of 1000 random combinations that have been manually tested, the following combination was found to be very close to the threshold of being more perceptive towards practical observations. However, there is a possibility of existence of more advantageous combinations cannot be ruled out.

$$M = X^1 Y^{-3} Z^3 \quad \text{for } \{a = 1, b = -3 \text{ and } c = 3)$$
$$L = X^0 Y^1 Z^0 \quad \text{for } \{a = 0, b = 1 \text{ and } c = 0)$$
$$T = X^0 Y^0 Z^1 \quad \text{for } \{a = 0, b = 0 \text{ and } c = 1)$$
$$K = X^1 Y^{-1} Z^1 \quad \text{for } \{a = 1, b = -1 \text{ and } c = 1)$$
$$A = X^{1/2} Y^2 Z^{-1} \quad \text{for } \{a = 1/2, b = 2 \text{ and } c = -1)$$

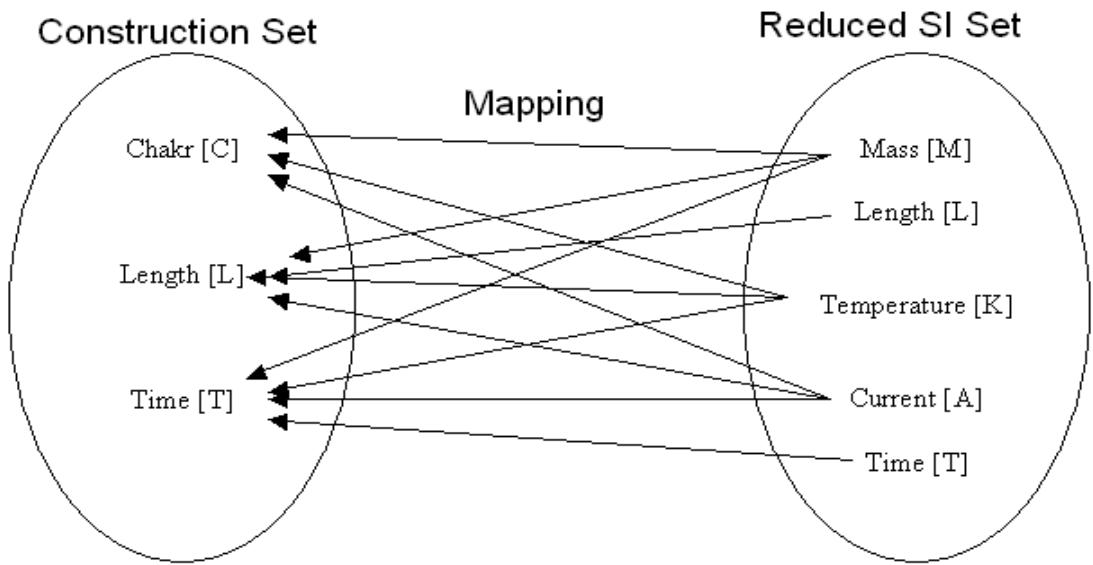

*Fig 1. Schematic View of Mapping done between Construction set and Reduced SI set*

Since $L = L = X^0Y^1Z^0$, we can interpret that $Y = L$. Similarly, $T = X^0Y^0Z^1$ can be interpreted as $Z = T$. Hence the new system is introducing only one unknown quantity "X". To make the new dimensional set sound more realistic, lets call this new base quantity as "Chakr" (Wheel) and represent it with letter "C". Hence the new dimensional system consists of Length [L], Time [T] and Chakr [C], lets call this new set as "Construction Set".

$$M = C^1 L^{-3} T^3$$
$$L = L^1$$
$$T = T^1$$
$$K = C^1 L^{-1} T^1$$
$$A = C^{1/2} L^2 T^{-1}$$

Substituting the values of {M, L, T, K and A} in the dimensional formulae of the derived quantities, we obtain

Velocity = $[L^1 T^{-1}] = [L^1 T^{-1}]$
Acceleration = $[L^1 T^{-2}] = [L^1 T^{-2}]$
Force = $[M^1 L^1 T^{-2}] = [C^1 L^{-2} T^1]$
Energy = $[M^1 L^2 T^{-2}] = [C^1 L^{-1} T^1]$
Universal Gravitational Constant = $[M^{-1} L^3 T^{-2}] = [C^{-1} L^6 T^{-5}]$
Charge = $[AT] = [C^{1/2} L^2]$
Voltage = $[M^1 L^2 T^{-3} A^{-1}] = [C^{1/2} L^{-3} T^1]$
Magnetic Pole Strength = $[A^1 L^1] = [C^{1/2} L^3 T^{-1}]$
Magnetic Moment = $[A^1 L^2] = [C^{1/2} L^4 T^{-1}]$
Intensity of magnetization = $[A^1 L^{-1}] = [C^{1/2} L^1 T^{-1}]$
Specific Heat = $[L^2 T^{-2} K^{-1}] = [C^{-1} L^3 T^{-3}]$
Thermal Capacity = $[M^1 L^2 T^{-1} K^{-1}] = [T]$
Permeability = $[M^1 L^1 T^{-2} A^{-2}] = [L^6 T^3]$
Permittivity = $[M^{-1} L^{-3} T^4 A^2] = [L^4 T^{-1}]$
Impedance = $[M^1 L^2 T^{-3} I^{-2}] = [L^{-5} T^2]$
[Check Appendix–I for other derived quantities]

The construction set is distinct, non-redundant and satisfies all the norms of existing literature. The next section throws light on the intuitive reasons favoring the construction set.

7. **INTUITIVE REASONS FAVORING THE CONSTRUCTION SET**

Here are few highlights of the construction set

1. One of the first remarkable changes brought by the Construction set is the change in the dimensional formula of Mass. There has been no reported discovery of any particle that has non-zero mass and has volume equal to zero. It can be argued that mass occupies volume. The new dimensional formula reflects this argument by incorporating $L^3$ (dimensional formula of volume) in its dimensional formula. Hence mass can be defined as existence of some derived quantity with dimensional formula $[C^1T^3]$ per unit volume. So far, practical evidence of neither Higgs Bosons nor Higgs field [12-17] has been found, there is possibility that mass might turn out to be a derived quantity as predicted by Construction set.

2. Permittivity and Permeability are vacuum constants and yet they have "M" (mass) included in their dimensional system as per SI units. Consider this thought example,

   > "suppose Alice exists in a isolated universe that contains no mass, in other words for Alice there is no such thing called as mass. Alice decides to measure vacuum constants i.e. permittivity and permeability. Will Alice include 'Mass' in the dimensional formulae of the vacuum constants?"

   The answer to the above thought example is "no" since Alice doesn't recognize something called as "mass". The new dimensional system accurately depicts this scenario unlike the older version with only length and time in their dimensional formulae. The new dimensional formulae of permittivity and permeability are $[L^4T^{-1}]$ and $[L^6T^3]$.

3. It is known that Vacuum has impedance, equal to 376.7 Ω. This is widely used in communication system as "free space impedance" [18]. Let us revisit the thought example that we considered previously,

   > "Now Alice decides to measure the free space impedance, will Alice includes "Mass" in the dimensional formula of impedance?" The answer is no, since Alice doesn't recognize something called as "mass"

   Again, the answer is no, since Alice doesn't recognize something called as 'mass'. If vacuum has impedance, then it is obvious that it should not have 'mass' in dimensional formula. The new dimensional system developed here rightly depicts this scenario. The new dimensional formula of "Impedance" is $[L^{-5}T^2]$.

4. One of the significant results obtained through this new dimensional system is that fact that the dimensional formula of "Energy" is equivalent to Temperature. It is known in literature that at increase in temperature corresponds to increase in rate of movements of atom in the system. This is also attributed as increase in the thermal energy of the system. In the fields of plasma physics and quantum electrodynamics, temperature is calculated in terms of electron volts and then converted into Kelvin by multiplying with suitable constant. From the above cases, it can be seen that Temperature is nothing but a form of energy and hence the dimensional formula of Temperature and Energy are equivalent. Rightly, the gas constant, which represents a mere statistical number, is a dimensionless quantity.

5. Considering the fact that "Length" and "Time" have retained their position as fundamental quantities and quantities such as "Mass", "Temperature" and "Current" are derived in terms of "Length", "Time" and "Chakr". The proposal of existence of unknown fundamental quantity "Chakr" is the most noteworthy concept of this new dimensional system. Potentially the concept of "Chakr" solves the trivial problem existing in the current physics. According to the standard model, fundamental particles obtain their masses upon interacting with the scalar background Higg's field, which happens to be in non zero ground state [13-17]. However there is an ambiguity about the existence of the Higg's field. The new dimensional formula of "mass" entirely changes the point of view regarding "mass". As per the new dimensional system, mass is a mere derived quantity.

6. Law of Conservation of Fundamental quantities of Construction Set: Any fundamental quantity of the Construction set can neither be created nor destroyed. This is a new insightful aspect of Construction Set unlike the older set of {M, L, T, A and K}. In the older set, we had the possibility such as "Mass can be totally destroyed to obtain energy and this energy can be converted into electrical energy or heat energy". The simplest of the argument that can be made here is that "if a particular quantity is fundamental, it can neither be created or destroyed, it exists for eternal".

7. In the older dimensional system, the presence of mass in the dimensional formula of Energy has resulted in constraints such as zero rest mass. It is to say that any physical entity that has energy and non-existent mass is attributed as "zero rest mass". The new dimensional system resolves this issue, as mass and energy are not directly related dimensionally.

8. As per the new dimensional system, the dimensional formula of derived quantities is more logical compared with previous one. Gravitational and Electrical forces are approximated with inverse square law. This effect is reflected in the dimensional formula of Force, which has [L²] in it.

9. In the quantum field theories, charge is treated as fundamental quantity whereas the dimensional system treats electric current as a fundamental quantity. It is startling to note that "charge" has "time" in its dimensional formula even though existence of charge is independent of time. As per the construction set, the dimensional formula of charge is independent of time and the dimensional formula of current is dependent on time.

10. With the ongoing conflict between quantum mechanics and relativity, the Construction set offers a new paradigm to unify both the theories. The construction set is perfectly compatible with the existing literature plus it offers a whole new dimensional quantity that can be defined as per the needs of unification of the two big theories [1-8].

**CONCLUSIONS**

Incorrect assumption of fundamental quantities can negatively affect the success of the theories. The incompatibility of the modern day theories is a possible example of incorrect choice of fundamental quantities. In this paper, a new set of fundamental quantities is described by removing the redundancy in the SI fundamental units. The New set, referred in the paper as "Construction Set" proposes a new fundamental quantity called as "Chakr" [C]. The Construction set includes three fundamental quantities namely, Length, Time and Chakr. All the other derived quantities are dimensionally described in terms of Length, Time and Chakr. The basic mapping between the set {Length, Time and Chakr} and {Mass, Length, Time, Current and Temperature} is evaluated after analyzing 1000 mappings generated by the Computer Program. The paper also describes the intuitive reasons supporting the new "Construction Set" as ideal set of fundamental quantities. The scope of future work includes studying the properties of Chakr and description of modern theories in terms of Construction set.

# Appendix – I

| Serial Number | Quantity | SI dimensions | Construction Set |
|---|---|---|---|
| 1 | Length | L | L |
| 2 | Time | T | T |
| 3 | Chakr | - | C |
| 4 | Energy | $ML^2T^{-2}$ | $CL^{-1}T$ |
| 5 | Area | $L^2$ | $L^2$ |
| 6 | Volume | $L^3$ | $L^3$ |
| 7 | Velocity | $LT^{-1}$ | $LT^{-1}$ |
| 8 | Acceleration | $LT^{-2}$ | $LT^{-2}$ |
| 9 | Force | $MLT^{-2}$ | $Cl^{-2}T$ |
| 10 | Mass | M | $CL^{-3}T^3$ |
| 11 | Mass density | $ML^{-3}$ | $CL^{-6}T^3$ |
| 12 | Electric current | A | $C^{1/2}L^2T^{-1}$ |
| 13 | Charge | AT | $C^{1/2}L^2$ |
| 14 | Temperature | K | $CL^{-1}T$ |
| 15 | Momentum | $MLT^{-1}$ | $CL^{-2}T^2$ |
| 16 | Impulse | $MLT^{-1}$ | $CL^{-2}T^2$ |
| 17 | Power | $ML^2T^{-3}$ | $CL^{-1}$ |
| 18 | Angular displacement | No dimensions | Dimensionless |
| 19 | Angular velocity | $T^{-1}$ | $T^{-1}$ |
| 20 | Angular acceleration | $T^{-2}$ | $T^{-2}$ |
| 21 | Angular momentum | $ML^2T^{-1}$ | $CL^{-1}T^2$ |
| 22 | Moment of Inertia | $ML^2$ | $CL^{-1}T^3$ |
| 23 | Frequency | $T^{-1}$ | $T^{-1}$ |
| 24 | Plank's constant | $ML^2T^{-1}$ | $CL^{-1}T^2$ |
| 25 | Coefficients of restitution | No dimension | Dimensionless |
| 26 | Force constant | $MT^{-2}$ | $CL^{-3}T$ |
| 27 | Stress | $ML^{-1}T^{-2}$ | $CL^{-4}T$ |
| 28 | Strain | No dimension | Dimensionless |
| 29 | Elastic modulii | $ML^{-1}T^{-2}$ | $CL^{-4}T$ |
| 30 | Poisson's ratio | No dimension | Dimensionless |
| 31 | Surface tension | $MT^{-2}$ | $CL^{-3}T$ |
| 32 | Coefficient of viscosity | $ML^{-1}T^{-1}$ | $CL^{-4}T^2$ |
| 33 | Velocity gradient | $T^{-1}$ | $T^{-1}$ |
| 34 | Universal gravitational constant | $M^{-1}L^3T^{-2}$ | $C^{-1}L^6T^{-5}$ |
| 35 | Heat | $ML^2T^{-2}$ | $CL^{-1}T$ |
| 36 | Coeff. Of thermal expansion | $K^{-1}$ | $C^{-1}LT^{-1}$ |
| 37 | Specific heat | $L^2T^{-2}K^{-1}$ | $C^{-1}L^3T^{-3}$ |
| 38 | Thermal capacity | $ML^2T^{-1}K^{-1}$ | $T$ |
| 39 | Gas constant | $ML^2T^{-2}K^{-1}$ | Dimensionless |
| 40 | Botlzmann constant and entropy | $ML^2T^{-2}K^{-1}$ | Dimensionless |
| 41 | Latent heat | $L^2T^{-2}$ | $L^2T^{-2}$ |
| 42 | Coeff. Of thermal conductivity | $MLT^{-3}K^{-1}$ | $L^{-1}T^{-1}$ |
| 43 | Stefan's constant | $MT^{-3}K^{-4}$ | $C^{-3}LT^{-4}$ |
| 44 | Magnetic pole strength | $AL$ | $C^{1/2}L^3T^{-1}$ |
| 45 | Magnetic moment | $AL^2$ | $C^{1/2}L^4T^{-1}$ |
| 46 | Flux density | $MT^{-2}A^{-1}$ | $C^{1/2}L^{-5}T^2$ |
| 47 | Magnetic flux | $ML^2T^{-2}A^{-1}$ | $C^{1/2}L^{-3}T^2$ |
| 48 | Intensity of magnetic field, Intensity of magnetization | $AL^{-1}$ | $C^{1/2}LT^{-1}$ |
| 49 | Permittivity | $M^{-1}L^{-3}T^4A^2$ | $L^4T^{-1}$ |
| 50 | Permeability | $MLT^{-2}A^{-2}$ | $L^{-6}T^3$ |
| 51 | Magnetic susceptibility | Nil | Dimensionless |
| 52 | Electric potential, E.M.F | $ML^2T^{-3}A^{-1}$ | $C^{1/2}L^{-3}T$ |
| 53 | Electric capacity | $M^{-1}L^{-2}T^4A^2$ | $L^5T^{-1}$ |
| 54 | Intensity of electric field | $MLT^{-3}A^{-1}$ | $C^{1/2}L^{-4}$ |
| 55 | Electric resistance | $ML^2T^{-3}A^{-2}$ | $L^{-5}T^2$ |
| 56 | Specific resistance | $ML^3T^{-3}A^{-2}$ | $L^{-4}T^2$ |
| 57 | Conductance | $M^{-1}L^{-2}T^3A^2$ | $L^5T^{-2}$ |
| 58 | Self inductance, mutual inductance | $ML^2T^{-2}A^{-2}$ | $L^{-5}T^3$ |
| 59 | Rydberg constant, wave number | $L^{-1}$ | $L^{-1}$ |
| 60 | Compressibility | $M^{-1}LT^2$ | $C^{-1}L^4T^{-1}$ |